\documentclass[twocolumn,aps,prl,showpacs,amsmath,amssymb,superscriptaddress]{revtex4-1}
\usepackage{epsfig}
\usepackage{graphicx}
\usepackage{dcolumn}
\usepackage{bm}
\usepackage[colorlinks,citecolor=blue]{hyperref}
\usepackage{multirow}

\begin{document}
\title{Fermion-induced Dynamical Critical Point}
\author{Shuai Yin}
\affiliation{School of Physics, Sun Yat-Sen University, Guangzhou 510275, China}

\author{Shao-Kai Jian}
\affiliation{Condensed Matter Theory Center, Department of Physics, University of Maryland, College Park, Maryland 20742, USA}

\date{\today}

\begin{abstract}
Dynamical phase transition (DPT) characterizes the abrupt change of dynamical properties in nonequilibrium quantum many-body systems.
It has been demonstrated that extra quantum fluctuating modes besides the conventional order parameter field can drastically change the properties of equilibrium phase transitions.
However, the counterpart phenomena in DPTs have rarely been explored.
Here, we study the DPT in the Dirac system after a sudden quench, and find that the fermion fluctuations can round a putative first-order DPT into a dynamical critical point, which is referred to as a fermion-induced dynamical critical point (FIDCP).
It is also a nonthermal critical point, in which the universal short-time scaling behavior emerges despite the system goes through a first-order transition after thermalization.
In the novel scenario of FIDCP, the quantum Yukawa coupling $g_q$ is indispensable for inducing the FIDCP albeit irrelevant in the infrared scale.
We call these variables {\it indispensable irrelevant scaling variables}.
Moreover, a dynamical tricritical point which separates the first-order DPT and the FIDCP is discovered by tuning this indispensable irrelevant scaling variable.
We further mention possible experimental realizations.
\end{abstract}
\maketitle
%
\emph{Introduction}.---Fathoming nonequilibrium dynamics of isolated quantum systems is one of the most important and challenging issues in modern statistical mechanics and condensed matter physics~\cite{Dz2010,Pol2011,Pol2016rev}.
On the one hand, these studies provide fundamental insights into how equilibrium thermodynamics emerges from a unitary time evolution~\cite{Deutsch1991,Srednicki1994,Rigol2008,Neill2016,Pol2016sci}.
For example, the eigenstate thermalization hypothesis attempts to build the gorgeous edifice of the statistical ensemble theory upon the cornerstone of the eigenstate properties of quantum many-body systems~\cite{Deutsch1991,Srednicki1994,Rigol2008}.
On the other hand, emergences of vibrant far-from-equilibrium phenomena in experiments are calling for new theoretical frameworks~\cite{Gring2012,Langen2013,Eigen,Berges2004,Mitra2018,Langen2016,Mori2018,Marcuzzi2013,Bertini2013,Mallayya2019}.
Among them the theory of dynamical phase transition (DPT) has attracted considerable attentions.
By analogy with the equilibrium phase transition, the DPT describes the abrupt change in dynamical properties in nonequilibrium systems~\cite{Cardy2006,Cardy2007,Werner2009,Biroli2010,Fabrizio2010,Demler2011,Werner2013,Werner20131,Biroli2013,Heyl2013,Chandran2013,Silva2015,Heyl2018, Zhang2017,Smale2019}.
It has been shown that the appearance of the DPT can lead to the universal short-time scaling behavior~\cite{Mitra2015,Mitra20151,Mitra2016,Marino2017,Swingle2019} similar to the critical initial slip in classical~\cite{Janssen,LiZB2015} and quantum dissipative systems~\cite{Yin2014,Schmalian2014,Schmalian2015}.

In equilibrium phase transitions, the importance of fluctuations cannot be overemphasized.
Long wave-length fluctuations are at the origin of scaling behaviors near second-order phase transitions, resulting in the concept of the universality class---one of organizational principles in condensed matter physics~\cite{Wilson}.
More strikingly, fluctuations can change the nature of the phase transition profoundly.
Coleman and Weinberg proposed a fluctuation-induced first-order phase transition by coupling the order parameter to a fluctuating gauge field~\cite{Weinberg}.
This found important applications in the context of phase transitions in the early universe and superconductors.
On the other hand, the theory of deconfined quantum critical point~\cite{Senthilsci2004,Senthilprb2004,Sandvik2007,Nogueira2007,Melko2008,Block2013,Lou2009,Pujari2013,Nahum2015A,Wang2015,Shao2016,Nahum2015B,Sato2017,Sreejith2019,Yao2019} takes the opposite track by showing that extra fluctuations from emergent degrees of freedom can soften the putative first-order phase transition~\cite{Landau} to be a continuous one.
Another example is the fermion-induced quantum critical point (FIQCP)~\cite{Li2017}, in which the extra fluctuations come from massless Dirac fermions.
It has been shown that both the Landau-de Gennes and the Landau-Devinshire first-order phase transitions can be rounded into continuous ones by fermion fluctuations~\cite{Scherer2016,Classen2017,Jian2017A,Jian2017B,Torres2018,Roy2019,Yin2020}.
Given these novel examples in equilibrium physics, equally important questions in the context of nonequilibrium physics arise: To what extent is the DPT affected by extra fluctuations?
Is the notion of continuous/discontinuous transition in the DPT the same as in the equilibrium phase transition?

\begin{figure}
 \centering
   \includegraphics[bb= 0 0 500 230, clip, scale=0.45]{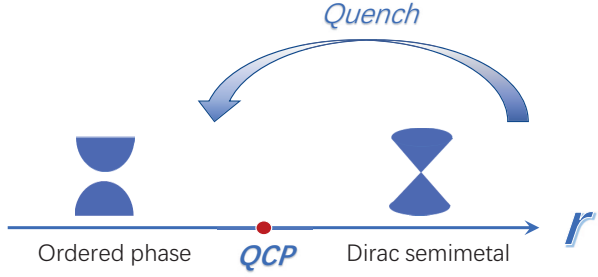}
   \caption{\label{quench} A schematic diagram of the quench protocol. The Dirac system hosts a Dirac semimetal phase, where the Dirac fermion is gapless and the bosonic order parameter is zero, and an ordered phase, where the Dirac fermion is gapped by a chiral mass determined by the non-zero order parameter. These two phases are separated by a quantum critical point. The system is initially prepared in the deep Dirac semimetal phase with a boson mass $\Omega$, then suddenly quenched into the vicinity of a dynamical phase transition (DPT) point. We show that the fermion fluctuation can change the first-order DPT into a dynamical critical point, which is referred to as a fermion-induced dynamical critical point.}
\end{figure}

According to Landau's paradigm~\cite{Landau}, the Landau-Devinshire model with negative quartic interactions leads to a discontinuous transition~\cite{Devonshire}, and putatively its quench dynamics also features a first-order DPT without any finite fixed point.
Although fermion fluctuations play no role at finite temperature in the equilibrium transitions~\cite{Stephanov1995,Wessel2016}, surprisingly, this is not true at nonequilibrium.
In this paper, we report a fermion-induced dynamical critical point (FIDCP) in the Dirac systems after a sudden quench, where we find that the putative first-order DPT can be driven into a continuous one by fermion fluctuations, i.e., the equilibrium Landau's paradigm is not longer valid at nonequilibrium.
This FIDCP corresponds to a dynamical chiral Ising fixed point~\cite{Swingle2019}, which is a nonthermal fixed point and controls the nonequilibrum scaling behavior near the FIDCP.
Moreover, we find that although the quantum Yuakwa coupling $g_q$ is irrelevant near the dynamical chiral Ising fixed point, it plays an indispensable role in bringing out the FIQCP, in contrast to the conventional irrelevant scaling variable which is negligible in equilibrium phase transitions.
We refer to this kind of scaling variable as the ``\textit{indispensable irrelevant scaling variable}".
Like the dangerously irrelevant parameters in the deconfined quantum critical point and the FIQCP~\cite{Shao2016,Jian2017B,Torres2018}, this indispensable irrelevant scaling variable is vital in determining the universal nonequilibrium behavior, and makes the notion of continuous/discontinuous transitions in nonequiblibrim physics distinct from the thermal ones.
Associated with the indispensable irrelevant scaling variable, a dynamical tricritical point (DTCP), which is the watershed between the first-order DPT and the FIDCP, is then discovered.
We also point out sharp physical consequences which should be within the experimental reach.

\emph{Model and quench protocol}.---We consider the quench dynamics in Dirac systems with $N$ flavors of two-component Dirac fermions, as sketched in Fig.~\ref{quench}.
The system is initially prepared in the semimetal phase with a boson mass $\Omega^2 > 0$, then at $t=0$ the boson mass is suddenly changed to be $r$ and the system evolves according to the postquench Hamiltonian.
The nonequilibrium dynamics is described by the generating function $Z={\rm Tr} [e^{i S_K}]$~\cite{Mitra2015,Mitra20151,Mitra2016,Marino2017}.
The Keldysh action $S_K\equiv i S_i+S_b$ therein consists of two parts, $S_i$ and $S_b$, corresponding, respectively, to the initial state and the postquench dynamics, where~\cite{Swingle2019}
\begin{eqnarray}
\begin{aligned}
S_i=&\frac{1}{2}\int_x\int_0^\infty d\tau \left[(\partial_\tau\phi)^2+(\nabla\phi)^2+\Omega^2\phi^2\right],\\
S_b=&\int_x\int_0^\infty dt [(\dot{\phi}_q\dot{\phi}_c-\nabla\phi_c\nabla\phi_q-r\phi_c\phi_q)\\
&-\frac{2u_c}{4!}\phi_c^3\phi_q-\frac{2u_q}{4!}\phi_q^3\phi_c+\Psi^\dag (i \partial_t+i \vec{\sigma}\cdot\nabla)\Psi_c\\
&-\frac{g_c}{\sqrt{2}}\phi_c\Psi^\dag\sigma_z\Psi-\frac{g_q}{\sqrt{2}}\phi_q\Psi^\dag\tau_x\sigma_z\Psi+...],
\label{eaction}
\end{aligned}
\end{eqnarray}
in which the subscripts $c$ and $q$ represent the classical and quantum parts of the action, respectively, in the Keldysh representation.
$\phi$ is the Ising boson field, and $\Psi\equiv(\psi_c,\psi_q)^T$ is the Dirac fermion field.
The summation over $N$ flavors is assumed.
$\int_x \equiv \int d^d x$, where $d$ is the spatial dimension, $\vec{\sigma}\equiv(\sigma_x,\sigma_y)$ is the Dirac matrix in two dimensions and can be generalized accordingly to higher dimensions, and $\tau$ acts on the Keldysh contour.
$u$ is the boson quartic coupling, $g>0$ is the Yukawa coupling, and the ellipses represent the higher-order terms to stabilize the system if necessary~\cite{supmat}.

In the equilibrium limit with $\Omega^2=r$, for $N=4$ and $d=2$ Eq.~(\ref{eaction}) with a positive $u$ describes the phase transition between the Dirac semimetal to charge density wave phases in graphene systems, while for $d=3$ it can describe the magnetic phase transition in Weyl/Dirac systems~\cite{Boyackreview}.
In addition, it has been demonstrated that when $u<0$ the putative bosonic Landau-Devonshire first-order phase transition~\cite{Devonshire} can be rounded by fermion fluctuations at zero temperature, giving rise to the type-II FIQCP~\cite{Yin2020}, which shares the same chiral universality class with  $u>0$~\cite{Boyackreview}.
Moreover, it was pointed out the thermal phase transition near the type-II FIQCP should be first order since the long-range fermion fluctuations, which play essential roles to soften the discontinuity in the first-order phase transition, is inhibited by the finite Matsubara-frequency gap proportional to the temperature~\cite{Yin2020}.

We will focus on the deep quench for $\Omega\gg\Lambda$ with $\Lambda^2$ being the UV momentum scale.
In this situation, the DPT is tuned by the renormalized boson mass,
\begin{eqnarray}
 r_{\rm eff}(t)&& = r+\frac{1}{2}\int d^d k D_K(t,t) \nonumber\\
 && -\frac{g_cg_q}{2}\int_0^t dt'\int d^d k {\rm Tr}[\tau_0\sigma_z G(t,t')\tau_x\sigma_zG(t',t)],
\label{mass}
\end{eqnarray}
in which $D_K(t,t')\equiv-i\langle \phi_c(t)\phi_c(t')\rangle$ is the boson Keldysh Green function and $G(t,t')\equiv-i\langle\Psi(t)\Psi^\dagger(t')\rangle$ is the fermion Green function.
For the deep quench, $D_K(t,t')\simeq-i\Omega[{\rm cos}\omega_k(t-t')-{\rm cos}\omega_k(t+t')]/\omega_k^2$~\cite{Mitra2015,Mitra20151,Mitra2016,Marino2017}, $\omega_k^2 \equiv {\vec k^2+r}$.
Note that the initial condition is contained in $D_K(t,t')$ and the second term in $D_K(t,t')$ breaks the time-translational symmetry explicitly.
By comparing $D_K(t,t')$ with the thermal Keldysh function, $D^{\rm (th)}_K(t,t') \simeq 2T {\rm cos}\omega_k(t-t')/\omega_k^2$, one finds that the $\Omega$ is similar to an effective temperature $T_{\rm eff}=\Omega/4$~\cite{Mitra2015,Mitra20151,Mitra2016,Marino2017}.
Although $r_{\rm eff}$ oscillates with a frequency proportional to $\Lambda$, the universal behavior of the DPT is contained in its time-independent part.
The DPT controlled by Eq.~(\ref{mass}) happens in a short-time stage before the thermalization time $t_{\rm th}$~\cite{Mitra2015,Mitra20151,Mitra2016,Marino2017}.
After $t_{\rm th}$, the secular terms with dissipation effects dominate and the system tends to the thermal sate~\cite{Mitraprl2011,Mitraprl2012}.

\begin{figure}[htb]
 \centering
   \includegraphics[bb= 0 120 330 250, clip, scale=0.7]{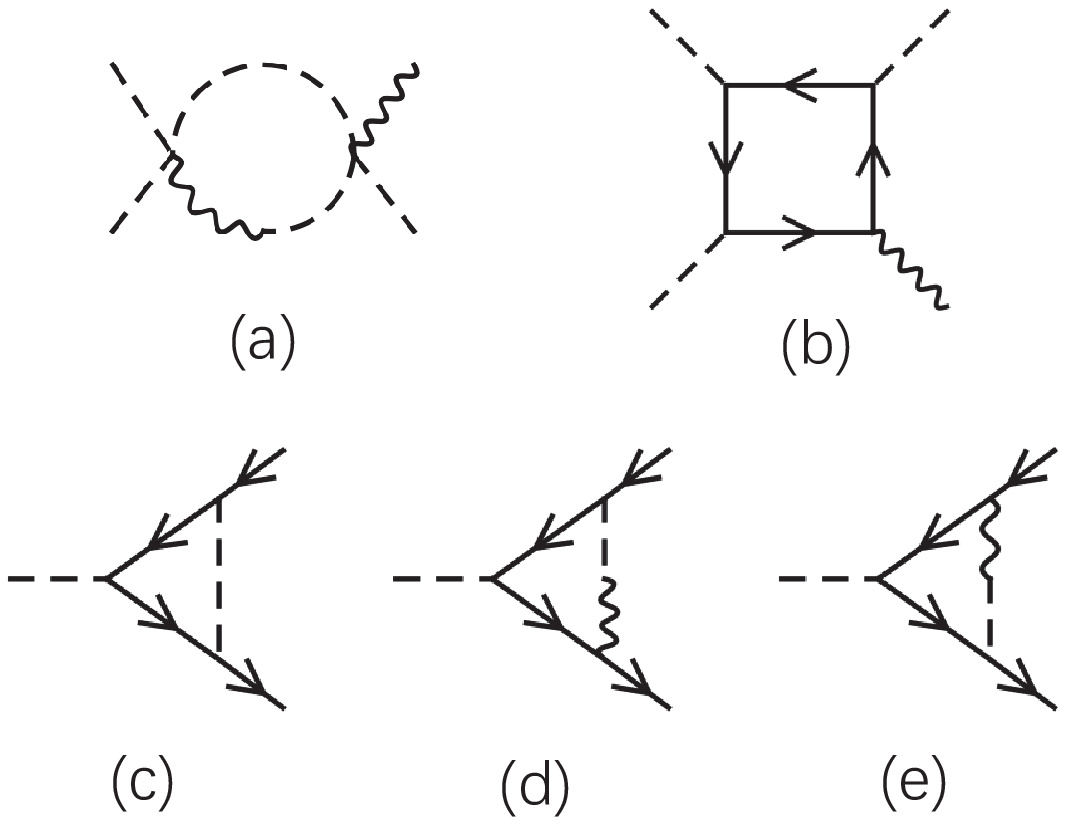}
   \caption{\label{Feynman} The Feynman diagrams for the one-loop corrections to $u_c$. The dashed line presents the classical boson field, the wiggly line presents the quantum bosonic field, and the solid line indicates the fermion field.}
\end{figure}

\emph{RG equations}.---To explore the DPT properties, we resort to the RG analyses.
By integrating out the momentum within the range $[\Lambda,\Lambda e^{-l}]$ ($l>0$ is the running parameter) for the inner line of the Feymann diagrams (some Feynman diagrams are shown in Fig.~\ref{Feynman} for illustration), and rescaling the couplings according to $k\rightarrow ke^l$, $u_c\rightarrow u_c e^{l(4-d-2\eta_b)}$, $u_c\rightarrow u_c e^{l(2-d-2\eta_b)}$, $g_c^2\rightarrow g_c^2 e^{l(4-d-\eta_b-2\eta_f)}$, and $g_q^2\rightarrow  g_q^2 e^{l(2-d-\eta_b-2\eta_f)}$, one obtains the following RG equations~\cite{Swingle2019,supmat},
\begin{eqnarray}
\frac{du_c}{dl}=&&(4-d-2 \eta_b)u_c-\frac{3}{8}u_c^2+6Ng_c^3g_q, \label{rguc}\\
\frac{du_q}{dl}=&&(2-d-2 \eta_b)u_c-\frac{3}{8}u_cu_q+6Ng_cg_q^3, \label{rguq}\\
\frac{dg_c^2}{dl}=&&(4-d-\eta_b-2 \eta_f)g_c^2-\frac{3}{8}g_c^4-\frac{3}{8}g_c^3g_q, \label{rggc}\\
\frac{dg_q^2}{dl}=&&(2-d-\eta_b-2 \eta_f)g_q^2-\frac{3}{8}g_c^2g_q^2-\frac{3}{8}g_cg_q^3, \label{rggq}
\end{eqnarray}
in which $\eta_b=Ng_cg_q/4$ and $\eta_f=g_c^2/12+g_cg_q/12$ are the anomalous dimensions for the boson and fermion fields, respectively.
The boson mass $r$ is relevant as a transition-tuning parameter, so we set $r$ to zero in Eqs.~(\ref{rguc}-\ref{rggq}) to describe scaling properties.
Although in the UV scale, $u_c=u_q$ and $g_c=g_q$, the classical part and the quantum part of the couplings have different dimensions for the deep quench case, since $\Omega$ is dimensionless, similar to the status of the temperature in classical phase transitions~\cite{Mitra2015,Mitra20151,Mitra2016,Marino2017}.
And we will see that the quantum Yukawa coupling plays a vital role in the FIDCP.

\emph{First-order DPT without Yukawa coupling}.---When $g_{c/q}=0$ and $u_{c/q}<0$ at the UV scale, Eqs~(\ref{rguc}) and (\ref{rguq}) show that in the IR scale $u_c$ tends to negative infinity.
It is quite different from the case for $u_c>0$, where a finite IR fixed point is reached~\cite{Swingle2019}.
Actually Eq.~(\ref{rguc}) is similar to the RG flow equation of the quartic boson coupling in the equilibrium $d$-dimensional Landau-Devonshire model~\cite{Devonshire,supmat}, indicating that the DPT is a first-order DPT~\cite{Devonshire}.
This consistency also exists between the $(d+1)$-dimensional dynamical fixed point and the $d$-dimensional Wilson-Fisher fixed point for the pure boson model~\cite{Mitra2015,Mitra20151,Mitra2016,Marino2017}.
The absence of the finite IR fixed point demonstrates that there is no self-similarity aging dynamics near this first-order DPT.

\emph{FIDCP with Yukawa coupling}.---Remarkably, the situation can be changed when the coupling to the Dirac fermion is introduced.
We will show that the gapless fluctuations of the Dirac fermion can trigger an emergent dynamical critical point, and as a result the universal dynamics governed by the long-wavelength modes near this FIDCP appears.
To see this, one can inspect Eq.~(\ref{rguc}).
The anticommutativity of the fermion fields leads to an additional minus sign in the fermionic loop diagram, as shown in Fig.~\ref{Feynman}~(b).
This makes the last term in Eq.~(\ref{rguc}) positive.
Accordingly, the last term makes an opposite contribution compared to the first two terms.
Heuristically, the direction of the RG flow of $u_c$ can be changed for large enough $g_{c/q}$.
\begin{figure}[htb]
  \centerline{\epsfig{file=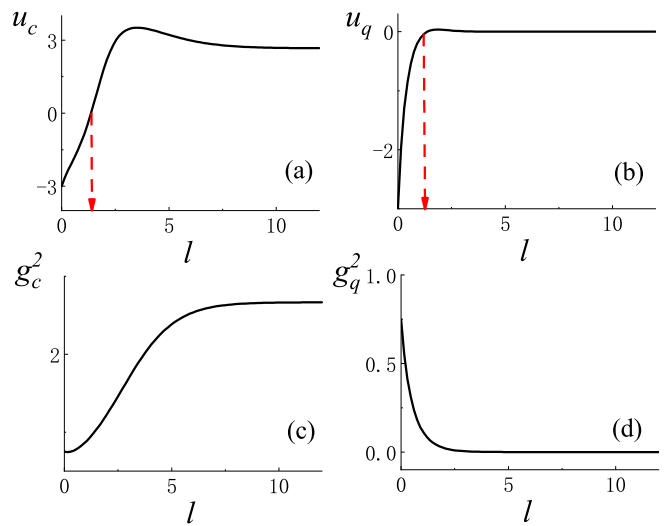,width=1\columnwidth}}
  \caption{\label{rgflowlam} For $N=2$ and $d=3$, the RG flows running from $l=0$ (UV) to $l\rightarrow \infty$ (IR) are shown in (a-d). The bare parameters are chosen as $u_c(0)=u_q(0)=-3$ and $g_c^2(0)=g^2_q(0)=0.75$. (a) shows that $u_c$ runs from a negative value to a positive one. (b) shows the $u_q$ also changes its sign and then tends to zero. The arrows in (a) and (b) denote positions of the sign changes for $u_c$ and $u_q$, respectively. $g_c$ tends to a finite fixed point as shown in (c), and $g_q$ tends to zero as shown in (d).
  }
\end{figure}

To quantitatively explore the FIDCP, we solve the RG equations~(\ref{rguc}-\ref{rggq}) explicitly by taking $N=2$ and $d=3$ as an example and show the results in Fig.~\ref{rgflowlam}.
From Fig.~\ref{rgflowlam}~(a) one finds that for a finite UV $g_{c/q}$, $u_c$ changes its sign from negative to positive at some intermediate scale, and then tends to an IR fixed point.
This intermediate scale decreases as $g_q(0)$ increases.
The appearance of the finite fixed point demonstrates that the low-frequency sector of the fluctuation is dominated by a critical point, rather than the first-order DPT.
Moreover, the massless boson correlation can induce a nontrivial Yukawa fixed point, as shown in Fig.~\ref{rgflowlam}~(c), via the one-loop correction to the Yukawa coupling.
The nonzero value of the $g_c$ indicates that the fixed point is a nonthermal fixed point, since it should be zero in the thermal critical point owing to the finite-Matsubara frequency gap.
Actually, this fixed point is just the dynamical chiral Ising fixed point reported previously~\cite{Swingle2019}.
The most remarkable phenomenon associated with the dynamical chiral Ising fixed point is the universal critical initial slip behavior, in which the boson order parameter $M$ change with time as $M\propto M_0t^{\theta}$ with $\theta$ being the critical initial slip exponent~\cite{Mitra2015,Mitra20151,Mitra2016,Marino2017,Swingle2019}.
Moreover, the fermion field has an anomalous dimension $\eta_f=(4-d)/12$~\cite{Swingle2019}.

Here we emphasize that the quantum part of the Yukawa coupling $g_q$ is a novel scaling variable.
Conventional irrelevant scaling variables are negligible near equilibrium phase transitions.
The situation, however, is quite different for $g_q$, though it tends to zero near the dynamical chiral Ising fixed point.
If $g_q$ were set to be zero in the UV scale, one finds from Eq.~(\ref{rguc}) that the fermion fluctuations do not participate in the postquench dynamics, resulting in a first-order DPT, as we discussed above.
It is the finite $g_q$, together with $g_c$, at the UV scale that brings fermion fluctuations into the boson potential, reverse the sign of $u_{c/q}$, and consequently results in the dynamical chiral Ising fixed point, and generates the universal critical initial slip behavior.
Thus, $g_q$ is an ``indispensable irrelevant scaling variable".

\begin{figure}[htb]
  \centerline{\epsfig{file=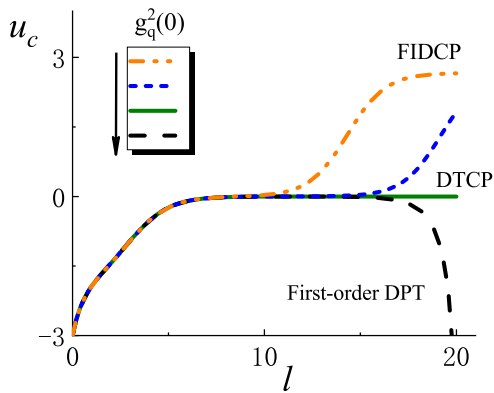,width=0.75\columnwidth}}
  \caption{\label{tric} For $N=2$ and $d=3$, a fixed point corresponding to a DTCP is determined by tuning $g_q$. Other parameters in the UV scale are chosen as $u_c(0)=u_q(0)=-3$ and $g^2_c(0)=g^2_q(0)$. At the DTCP, $g^2_{q\rm tr}(0)\simeq0.726624313854$. $u_c$ is relevant at the DTCP and its fixed point value is zero. $g_c$ tends to a finite value, indicating this is a nonthermal fixed point. Both $u_q$ and $g_q$ are irrelevant in the IR limit.
 }
\end{figure}

\emph{Dynamical tricritical point}.---Appearances of unusual irrelevant scaling variables can reshape critical properties in both the deconfined quantum critical point and the FIQCP~\cite{Shao2016,Jian2017B,Torres2018}.
Here we show in Fig.~\ref{tric} that the indispensable irrelevant scaling variable $g_q$ can manoeuvre a dynamical tricritical point (DTCP), though it is still irrelevant near this DTCP.
This DTCP appears at $g_{q \rm tr}(0)$ when other parameters are fixed in UV scale. And only for $g_q(0)>g_{q \rm tr}(0)$, can the FIDCP arise.
At the fixed point of this DTCP, $u_{c \rm tr}^\ast=0$.
Besides $r$ whose scaling is relevant, $r\propto re^{2l}$, Fig.~\ref{tric} shows that $u_c$ is the other relevant direction near the DTCP.
When $g_q(0)$ is close to $g_{q \rm tr}(0)$, $u_c$ lingers over a plateau for a period of scale, then tends to negative infinity or the dynamical chiral fixed point depending on whether $g_q(0) < g_{q \rm tr}(0)$ or $g_q(0) > g_{q \rm tr}(0)$.
From Eq.~(\ref{rguc}) one finds that $u_c$ deviates from $u_{c \rm tr}^\ast$ by $u_c\propto u_c e^{l}$.
Surprisingly, $g_c$ tends to a finite value at DTCP, indicating that this fixed point is also a nonthermal fixed point.
Although the fixed point value of $g_c$ is equal to that at the dynamical chiral Ising fixed point, this is a one-loop result, and the higher-order boson field renormalization can distinguish them.
Moreover, the larger the fermion flavor number $N$ is, the smaller $g_q(0)$ is needed to induced the FIDCP~\cite{supmat}, since more fermion fluctuations can be taken into account, which can also be found from Eq.~(\ref{rguc}).

\emph{Discussion}.---The results obtained above provide several sharp experimental signatures.
To see this, we compare different cases.
(a) When $u>0$, the universal behaviors not only exist in the short-time stage, but also manifest themselves after thermalization.
According to the eigenstate thermalization hypothesis, these scaling behaviors in the thermalization stage is just the classical phase transition governed by the Wilson-Fisher fixed point~\cite{Srednicki2015,Stephanov1995,Wessel2016}.
(b) When $u<0$ and $g<g_{\rm tr}$ is small, the quench dynamics does not show any universal scaling behavior in any stage after the quench, since both the DPT and the thermal phase transition are first order.
(c) When $u<0$ and $g>g_{\rm tr}$ is large enough to bring out the FIDCP, universal scaling behaviors emerge in the short-time relaxation stage before the equilibration $t < t_{\rm th}$.
After the equilibration $t \gg t_{\rm th}$ in the thermal region, the Masubara frequency of fermion modes opens a gap proportional to the effective temperature $T_{\rm eff}\gg \Lambda^2$.
Thus the mechanism of the FIDCP is interdicted, and consequently the long-time thermal region exhibits no universal scaling properties.

Recently, the Dirac systems with tunable interactions are realized in cold atom systems~\cite{Greif2015,Bloch2017}.
In addition, manipulation and detection of nonequilibrium dynamics have been realized in various systems~\cite{Weiss2006,Hofferberth2007}.
In particular, short-time scaling behaviors were found in recent experiments~\cite{Erne,Oberthaler,Nicklas}.
Accordingly, the FIDCP studied here appears within the experimental reach.

\emph{Summary and outlook}.---In summary, we have studied dynamical phase transitions in Dirac systems and reported a fermion-induced dynamical critical point (FIDCP).
We have showed that a first-order DPT for the pure boson model can be rounded by the fermion fluctuations into a dynamical critical point, corresponding to the dynamical chiral Ising fixed point in our case.
In the novel scenario of FIDCP as we have discussed, the quantum Yukawa coupling is an indispensable irrelevant scaling variable, which plays a crucial role in inducing the FIDCP though it is irrelevant near the critical point.
The existence of indispensable irrelevant scaling variables makes the notion of continuous/discontinuous transitions in dynamical phase transitions different from that in equilibrium ones.
Furthermore, a DTCP associated with the indispensable irrelevant scaling variable, i.e., $g_q$, has been identified.
It will be interesting to explore this new tricritical point, which we leave as further work.

Our paper not only gives the first example that the fluctuation can change the order of nonequilibrium dynamical phase transition, but also provides experimental criteria to detect it. Our results can be generalized in to the Dirac/Weyl systems with more boson field components in $d=3$ and the Ising and $XY$ cases in $d=2$.
Moreover, our results should also be applicable in itinerant electronic systems with finite Fermi surfaces.
In the equilibrium case, the tendency to turn the first-order transition into a continuous one was found in these systems~\cite{Jakubczyk,Jakubczyk2009}, so it is instructive to explore the nonequilibrium dynamics therein.

{\it Acknowledgement.}---We wish to thank F. Zhong and G.-Y. Huang for their helpful discussions. S.Y. is supported by the startup grant (No. 74130-18841229) at Sun Yat-Sen University. S.-K.J. is supported by the Simons Foundation through the It from Qubit Collaboration. SKJ and SY have equal contributions.

\begin{widetext}
\clearpage

\section{Supplemental Material}
\renewcommand{\theequation}{S\arabic{equation}}
\setcounter{equation}{0}
\renewcommand{\thefigure}{S\arabic{figure}}
\setcounter{figure}{0}
\renewcommand{\thetable}{S\arabic{table}}
\setcounter{table}{0}

\subsection{A. The Green functions}
To obtain the RG equations, the Gaussian Green's functions for both fermion and boson fields are needed. For a general initial boson mass $\Omega$, the Green functions for the boson fields reads~\cite{Mitra2015S,Mitra20151S,Mitra2016S,Marino2017S,Swingle2019S}
\begin{eqnarray}
D_R(t-t')=&&-\Theta(t-t')\frac{{\rm sin} \omega_k (t-t')}{\omega_k}, \label{tf1}\\
D_K(t,t')=&&-i \frac{1}{\omega_k}[K_+ {\rm cos} \omega_k (t-t')+K_- {\rm cos} \omega_k (t+t')], \label{tf2}
\end{eqnarray}
in which $D_R$ is the retarded Green function, $D_K$ is the Keldysh Green function $\omega_k=\sqrt{k^2+r}$ and $K_{\pm}=\frac{1}{2}[\frac{\omega_k}{\omega_{0k}}\pm \frac{\omega_{0k}}{\omega_{k}}]$ with $\omega_{0k}$ being $\omega_{0k}=\sqrt{k^2+\Omega}$. In the deep quench limit, in which $\Omega\gg \Lambda^2$, Eq.~(\ref{tf2}) can be simplified as
\begin{equation}
D_K(t,t')=-i \frac{\Omega}{\omega_k^2}[{\rm cos} \omega_k (t-t')- {\rm cos} \omega_k (t+t')]. \label{tf3}
\end{equation}
In addition, the Green functions for the fermion fields are~\cite{Swingle2019S}
\begin{eqnarray}
G_R(t-t')=&&-i\Theta(t-t')[e^{-ik(t-t')}P_+(k)+e^{ik(t-t')}P_-(k)], \label{tf4}\\
G_K(t,t')=&&-i [e^{-ik(t-t')}P_+(k)-e^{ik(t-t')}P_-(k)], \label{tf5}
\end{eqnarray}
in which $k=\sqrt{k_x^2+k_y^2}$ and $P_{\pm}=\frac{1}{2}(1\pm \hat{k}\cdot \vec{\sigma})$ with $\hat{k}=(k_x,k_y)/k$. Equations~(\ref{tf4}-\ref{tf5}) can be readily generalized to higher dimensions by taking more momentum components into account.

\subsection{B. FIDCP in $d=2$}
In this section, we show the FIDCP in $d=2$. By solving the RG equations (3-6) in the main text, we show that in two dimension the fermion fluctuations can also round the first-order DPT into a dynamical critical point, similar to the case in $d=3$. The results are plotted in Fig.~\ref{rgflowlam2}. In Fig.~\ref{rgflowlam2}, $u_{c/q}(0)$ are chosen to be identical to those in Fig.~3 in the main text, but $g_{c/q}^2(0)$ is smaller than that for $d=3$. This indicates that in two dimension, the effects induced by fluctuations are more apparent.

In addition, the DTCP for $d=2$, which is realized by tuning $g_q$, is also found as shown in Fig.~\ref{tric2d}. For the same other parameters, $g_{g \rm tr}$ is smaller than its counterpart in $d=3$.

\subsection{C. Tricritical point for different $N_f$}
In this section, we study the dependence of the $g_{q \rm tr}(0)$ on $N$. We take the case for $d=3$ as an example. Figure~\ref{gtrnf} shows the results. From Fig.~\ref{gtrnf} one finds that the tricritical point $g_{q \rm tr}^4(0)$ decrease as $N$ increases. By power fitting, one finds that $g_{q \rm tr}(0)$ satisfies $g^4_{q \rm tr}(0) \propto1/N$ approximatively. To see the reason, one can inspect Eq.~(3), from which one finds right hand side of Eq.~(3) changes its sign when
\begin{equation}
g^4(0)>\frac{3u_c(0)[6u_c(0)-8(4-d)]}{48N}.
\label{gtrvalue}
\end{equation}
Similar phenomena were found in the type-II fermion-induced quantum critical point~\cite{Yin2020S}.

\subsection{D. FIDCP with the higher-order boson coupling}
In the main text, we keep the terms which make leading contributions in the UV and IR scales near the transition point. When $u<0$, at least one positive higher-order boson coupling is needed to stabilize the system. In this section, we show that our main results are not altered by the higher-order boson couplings. Concretely, we assume that the sixth-order boson coupling is positive and higher-order terms are neglected. By casting this term into the closed time path integral, one obtain an additional term, $\int_0^\infty dt \int d^dx [-\frac{v_c}{6!} \frac{3}{2}\phi_c^5\phi_q-\frac{v_q}{6!}\frac{3}{2} \phi_q^5\phi_c-\frac{v_i}{6!} 5\phi_q^3\phi_c^3]$ in Eq.~(1).  Physically, the coupling $v(0)$ should be far smaller than $|u(0)|$ and $g(0)$ because it describes three-body collision processes. In the deep quench case, $v_q$ scales as $v_q\sim v_q(0) e^{l (2-2d-3 \eta_b)}$ and $v_i$ scales as $v_i\sim v_i(0) e^{l (4-2d-3\eta_b)}$. Both of them are less relevant than $u_q$ and play ignorable roles. We then obtain the one-loop RG equations by taking into account the contribution from $v_c$ as
\begin{eqnarray}
\frac{du_c}{dl}=&&(4-d-2 \eta_b)u_c-\frac{3}{8}u_c^2+6Ng_c^3g_q+\frac{1}{8}v_c, \label{rgucs}\\
\frac{dv_c}{dl}=&&(6-2 d-3 \eta_b)u_c+\frac{15}{8}u_c^3-\frac{15}{4}u_c v_c+\frac{345}{4}Ng_c^5g_q, \label{rgucs}\\
\frac{dg_c^2}{dl}=&&(4-d-\eta_b-2 \eta_f)g_c^2-\frac{3}{8}g_c^4-\frac{3}{8}g_c^3g_q, \label{rggcs}\\
\frac{dg_q^2}{dl}=&&(2-d-\eta_b-2 \eta_f)g_q^2-\frac{3}{8}g_c^2g_q^2-\frac{3}{8}g_cg_q^3. \label{rggqs}
\end{eqnarray}
By solving Eqs.~(\ref{rgucs}-\ref{rggqs}), one finds two nontrivial fixed points. One is $(u_c,g^2_c,v_c)=(\frac{8 \varepsilon}{3},\frac{8 \varepsilon}{3},O(\varepsilon^2))$. This is the dynamical chiral Ising fixed point~\cite{Swingle2019S}. The other is $(u_c,g^2_c,v_c)=(-\frac{16}{25}+\frac{88 \varepsilon}{75},\frac{8 \varepsilon}{3},\frac{768+384 \varepsilon}{625})$. This is the dynamical tricritical point with $u_c$ as its second relevant direction as we discussed in the main text. In Fig.~\ref{ucvc}, we show that the FIDCP can arise even with a finite $v_c(0)$.

\begin{figure}[htbp]
  \centerline{\epsfig{file=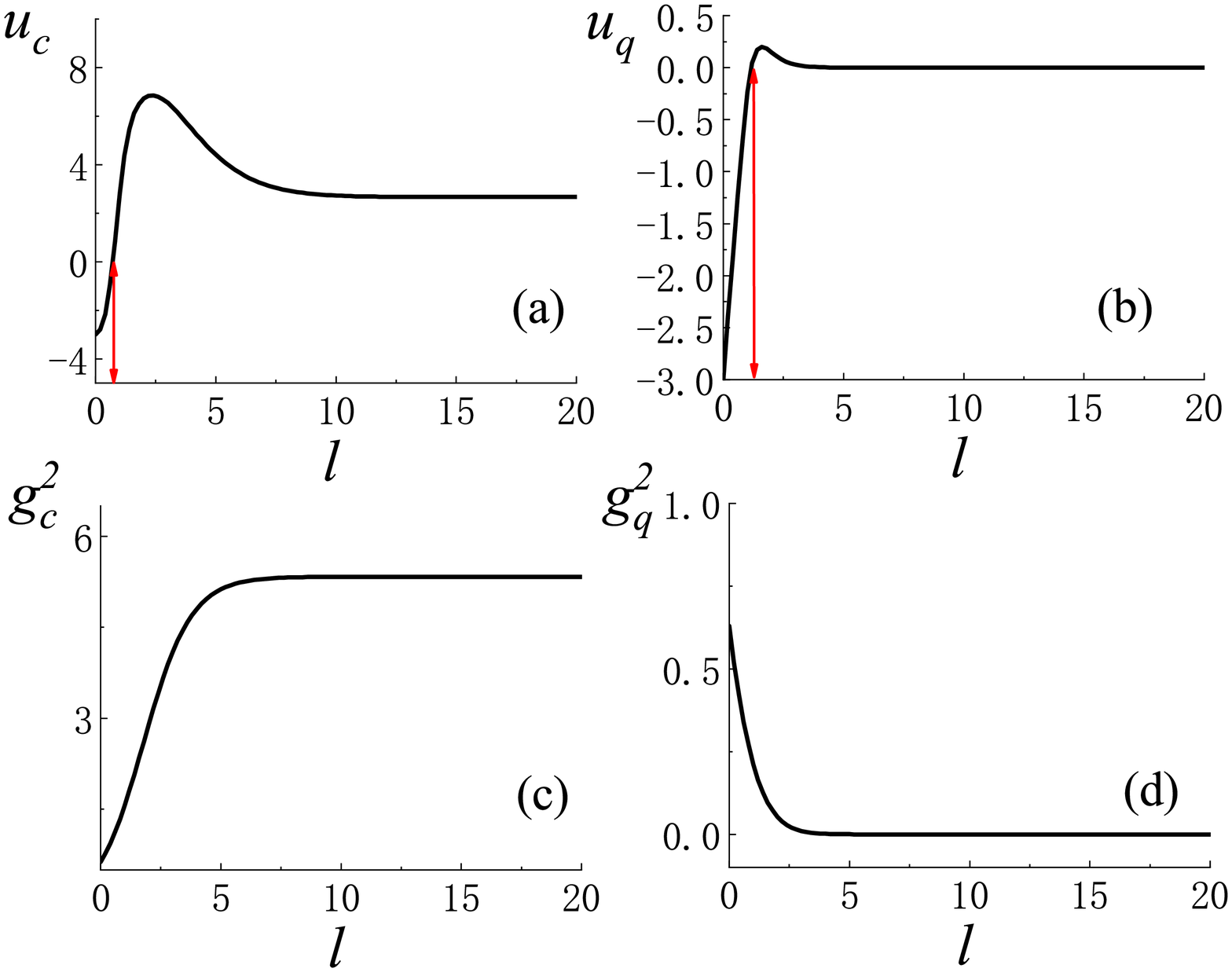,width=0.65\columnwidth}}
  \caption{\label{rgflowlam2} For $N=2$ and $d=2$, the RG flows run from $l=0$ (UV) to $l\rightarrow \infty$ (IR) are shown in (a-d). The bare parameters are chosen as $u_c(0)=u_q(0)=-3$ and $g_c^2(0)=g^2_q(0)=0.7$. (a) shows that $u_c$ runs from a negative value to a positive one. (b) shows the $u_q$ also changes its sign and then tends to zero. The arrows in (a) and (b) denote positions of the sign changing for $u_c$ and $u_q$, respectively. $g_c$ tends to a finite fixed point as shown in (c), and $g_q$ tends to zero as shown in (d).
  }
\end{figure}

\begin{figure}[htbp]
  \centerline{\epsfig{file=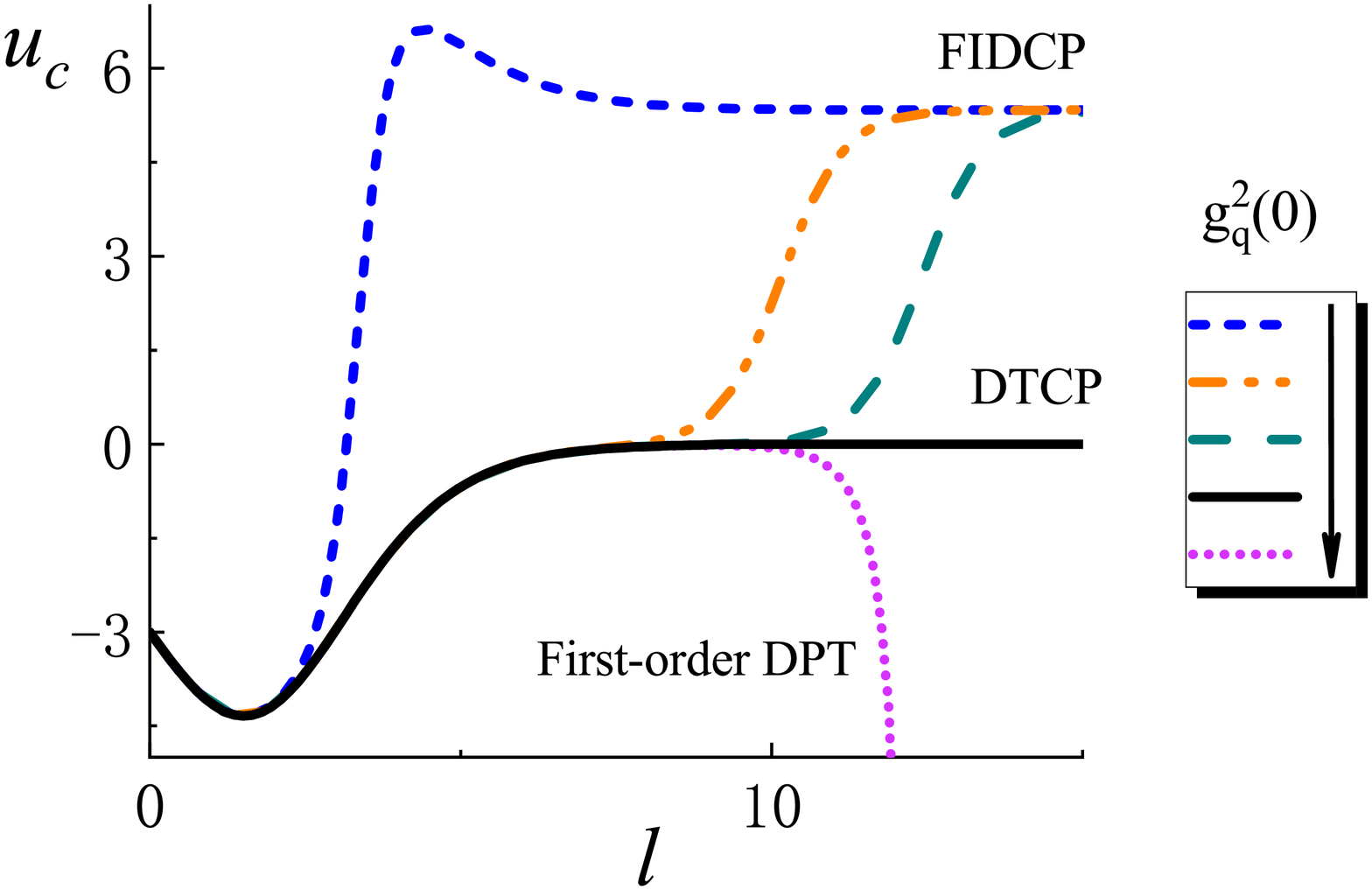,width=0.6\columnwidth}}
  \caption{\label{tric2d} For $N=2$ and $d=3$, a fixed point corresponding to a DTCP is determined by tuning $g_q$. Other parameters in the UV scale are chosen as $u_c(0)=u_q(0)=-3$ and $g^2_c(0)=g^2_q(0)$. At the DTCP, $g^2_{q\rm tr}(0)\simeq0.6991382539535241$. $u_c$ is relevant at the DTCP and its fixed point value is zero. $g_c$ tends to a finite value, indicating this is a nonthermal fixed point. Both $u_q$ and $g_q$ are irrelevant in the IR limit.
 }
\end{figure}

\begin{figure}[htbp]
  \centerline{\epsfig{file=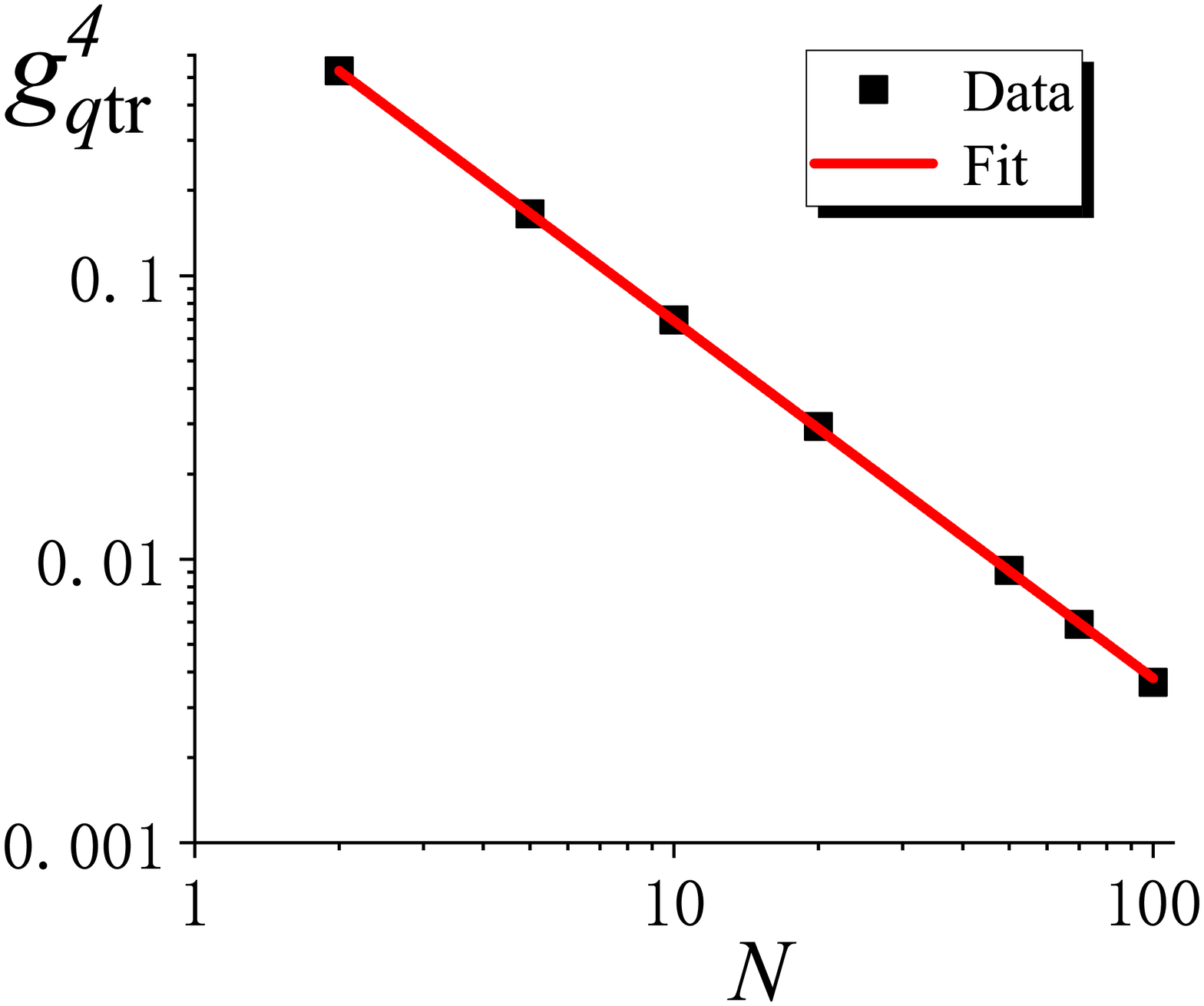,width=0.6\columnwidth}}
  \caption{\label{gtrnf} Curves of the tricritical point $g^4_{\rm tr}$ versus $N$. The boson quartic coupling is chosen as are chosen as $u(0)=-3$. Double logarithmic scales are used. Power fitting shows that $g^4_{\rm tr}\propto 1/N_f^{1.12}$.
  }
\end{figure}

\begin{figure}[htbp]
  \centerline{\epsfig{file=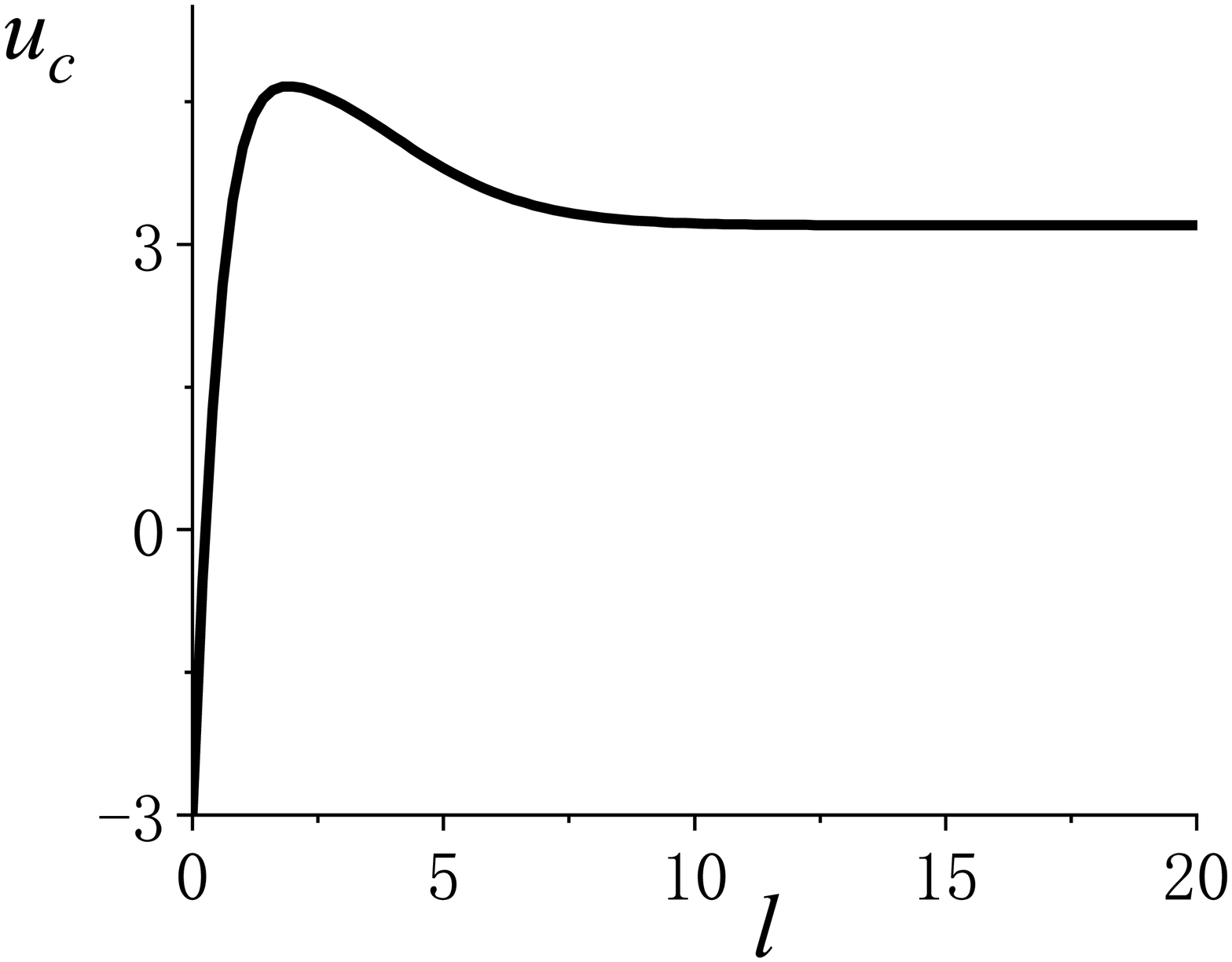,width=0.6\columnwidth}}
  \caption{\label{ucvc} For $N=2$ and $d=3$, the RG flow for $u_c$ run from $l=0$ (UV) to $l\rightarrow \infty$ (IR). The bare parameters are chosen as $u_c(0)=-3$, $v_c(0)=0.1$ and $g_c^2(0)=g^2_q(0)=1.2$. $u_c$ changes its sign and tends to the dynamical chiral Ising fixed point in the IR scale. For $g_{c/q}^2(0)<g_{c/q \rm tr}^2(0)\simeq 0.9$, $u_c$ tends to the negative infinity.
  }
\end{figure}

\end{widetext}


\begin{thebibliography}{99}
\bibitem{Dz2010} J. Dziarmaga, Adv. Phys. {\bf 59}, 1063 (2010).
\bibitem{Pol2011} A. Polkovnikov, K. Sengupta, A. Silva, and M. Vengalattore, Rev. Mod. Phys. {\bf 83}, 863 (2011).
\bibitem{Pol2016rev} L. D'Alessio, Y. Kafri, A. Polkovnikov, and M. Rigol, Adv. Phys., {\bf 65}, 239 (2016).
\bibitem{Deutsch1991} J. M. Deutsch, Phys. Rev. A {\bf 43}, 2046 (1991).
\bibitem{Srednicki1994} M. Srednicki, Phys. Rev. E {\bf 50}, 888 (1994).
\bibitem{Rigol2008} M. Rigol, V. Dunjko, and M. Olshanii, Nature (London) {\bf 452}, 854 (2008).
\bibitem{Neill2016} C. Neill, P. Roushan, M. Fang, Y. Chen, M. Kolodrubetz, Z. Chen, A. Megrant, R. Barends, B. Campbell, B. Chiaro, A. Dunsworth, E. Jeffrey, J. Kelly, J. Mutus, P. J. J. O'Malley, C. Quintana, D. Sank, A. Vainsencher, J. Wenner, T. C. White, A. Polkovnikov, and J. M. Martinis, Nat. Phys. {\bf 12}, 1037 (2016).

\bibitem{Pol2016sci} A. Polkovnikov and D. Sels, Science, {\bf 353}, 752, (2016).

\bibitem{Gring2012} M. Gring, M. Kuhnert, T. Langen, T. Kitagawa, B. Rauer, M. Schreitl, I. Mazets, D. A. Smith, E. Demler, and J. Schmiedmayer, Science {\bf 337}, 1318 (2012).
\bibitem{Langen2013} T. Langen, R. Geiger, M. Kuhnert, B. Rauer, and J. Schmiedmayer, Nat. Phys. {\bf 9}, 640 (2013).
\bibitem{Eigen} C. Eigen, J. A. P. Glidden, R. Lopes, E. A. Cornell, R. P. Smith, and Z. Hadzibabic, Nature (London) {\bf 563}, 221 (2018).
\bibitem{Berges2004} J. Berges, Sz. Bors\'anyi, and C. Wetterich, Phys. Rev. Lett. {\bf 93}, 142002 (2004).
\bibitem{Mitra2018} A. Mitra, Annu. Rev. Condens. Matter Phys. {\bf 9}, 245 (2018).
\bibitem{Langen2016} T. Langen, T. Gasenzer, and J. Schmiedmayer, J. Stat. Mech. (2016) 064009.
\bibitem{Mori2018} T. Mori, T. N Ikeda, E. Kaminishi, and M. Ueda, J. Phys. B {\bf 51}, 112001 (2018).
\bibitem{Marcuzzi2013} M. Marcuzzi, J. Marino, A. Gambassi, and A. Silva, Phys. Rev. Lett. {\bf 111}, 197203 (2013).
\bibitem{Bertini2013} B. Bertini, F. H. L. Essler, S. Groha, and N. J. Robinson, Phys. Rev. Lett. {\bf 115}, 180601 (2015).
\bibitem{Mallayya2019} K. Mallayya, M. Rigol, and W. De Roeck, Phys. Rev. X {\bf 9}, 021027 (2019).
\bibitem{Cardy2006} P. Calabrese and J. Cardy, Phys. Rev. Lett. {\bf 96}, 136801 (2006).
\bibitem{Cardy2007} P. Calabrese and J. Cardy, J. Stat. Mech. (2007) P06008.
\bibitem{Werner2009} M. Eckstein, M. Kollar, and P. Werner, Phys. Rev. Lett. {\bf 103}, 056403 (2009).
\bibitem{Biroli2010} B. Sciolla and G. Biroli, Phys. Rev. Lett. {\bf 105}, 220401 (2010).
\bibitem{Fabrizio2010} M. Schir\'{o} and M. Fabrizio, Phys. Rev. Lett. {\bf 105}, 076401 (2010).
\bibitem{Demler2011} T. Kitagawa, A. Imambekov, J. Schmiedmayer, and E. Demler, New J. Phys. {\bf 13}, 073018 (2011).
\bibitem{Werner2013} N. Tsuji and P. Werner, Phys. Rev. B {\bf 88}, 165115 (2013).
\bibitem{Werner20131} N. Tsuji, M. Eckstein, and P. Werner, Phys. Rev. Lett. {\bf 110}, 136404 (2013).
\bibitem{Biroli2013} B. Sciolla and G. Biroli, Phys. Rev. B {\bf 88}, 201110(R) (2013).
\bibitem{Heyl2013} M. Heyl, A. Polkovnikov, and S. Kehrein, Phys. Rev. Lett. {\bf 110}, 135704 (2013).
\bibitem{Chandran2013} A. Chandran, A. Nanduri, S. S. Gubser, and S. L. Sondhi, Phys. Rev. B {\bf 88}, 024306 (2013).
\bibitem{Silva2015} P. Smacchia, M. Knap, E. Demler, and A. Silva, Phys. Rev. B {\bf 91}, 205136 (2015).
\bibitem{Heyl2018} B. \v{Z}unkovi\v{c}, M. Heyl, M. Knap, and A. Silva, Phys. Rev. Lett. {\bf 120}, 130601 (2018).
\bibitem{Zhang2017} J. Zhang, G. Pagano, P. W. Hess, A. Kyprianidis, P. Becker, H. Kaplan, A. V. Gorshkov, Z.-X. Gong, and C. Monroe, Nature, {\bf 551}, 601 (2017).
\bibitem{Smale2019} S. Smale, P. He, B. A. Olsen, K. G. Jackson, H. Sharum, S. Trotzky, J. Marino, A. M. Rey, and J. H. Thywissen, Science Advances, {\bf 5}, eaax1568 (2019).
\bibitem{Mitra2015} A. Chiocchetta, M. Tavora, A. Gambassi, and A. Mitra, Phys. Rev. B {\bf 91}, 220302(R) (2015).
\bibitem{Mitra20151} A. Maraga, A. Chiocchetta, A. Mitra, and A. Gambassi, Phys. Rev. E {\bf 92}, 042151 (2015).
\bibitem{Mitra2016} A. Chiocchetta, M. Tavora, A. Gambassi, and A. Mitra, Phys. Rev. B {\bf 94}, 134311 (2016).
\bibitem{Marino2017} A. Chiocchetta, A. Gambassi, S. Diehl, and J. Marino, Phys. Rev. Lett. {\bf 118}, 135701 (2017).
\bibitem{Swingle2019} S.-K. Jian, S. Yin, and B. Swingle, Phys. Rev. Lett. {\bf 123}, 170606 (2019).

\bibitem{Janssen} H. K. Janssen, B. Schaub, and B. Schmittmann, Z. Phys. B {\bf 73}, 539 (1989).
\bibitem{LiZB2015} Z. B. Li, L. Sch\"ulke, and B. Zheng, Phys. Rev. Lett. {\bf 74}, 3396 (1995).
\bibitem{Yin2014} S. Yin, P. Mai, and F. Zhong, Phys. Rev. B {\bf 89}, 144115 (2014).
\bibitem{Schmalian2014} P. Gagel, P. P. Orth, and J. Schmalian, Phys. Rev. Lett. {\bf 113}, 220401 (2014).
\bibitem{Schmalian2015} P. Gagel, P. P. Orth, and J. Schmalian, Phys. Rev. B {\bf 92}, 115121 (2015).
\bibitem{Wilson} K. G. Wilson and J. Kogut, Phys. Rep {\bf 12}, 75 (1974).
\bibitem{Weinberg} S. Coleman and E. Weinberg, Phys. Rev. D {\bf 7}, 1888 (1973).



\bibitem{Senthilsci2004} T. Senthil, A. Vishwanath, L. Balents, S. Sachdev, and M. P. A. Fisher, Science {\bf 303}, 1490 (2004).
\bibitem{Senthilprb2004} T. Senthil, L. Balents, S. Sachdev, A. Vishwanath, and M. P. A. Fisher, Phys. Rev. B {\bf 70}, 144407 (2004).
\bibitem{Sandvik2007} A. W. Sandvik, Phys. Rev. Lett. {\bf 98}, 227202 (2007).
\bibitem{Nogueira2007} F. S. Nogueira, S. Kragset, and A. Sudb{\o}, Phys. Rev. B {\bf 76}, 220403(R) (2007).
\bibitem{Melko2008} R. G. Melko and R. K. Kaul, Phys. Rev. Lett. {\bf 100}, 017203 (2008).
\bibitem{Block2013} M. S. Block, R. G. Melko, and R. K. Kaul, Phys. Rev. Lett. {\bf 111}, 137202 (2013).
\bibitem{Lou2009} J. Lou, A. W. Sandvik, and N. Kawashima, Phys. Rev. B {\bf 80}, 180414(R) (2009).
\bibitem{Pujari2013} S. Pujari, K. Damle, and F. Alet, Phys. Rev. Lett. {\bf 111}, 087203 (2013).
\bibitem{Nahum2015A} A. Nahum, J. T. Chalker, P. Serna, M. Ortuno, and A. M. Somoza, Phys. Rev. X {\bf 5}, 041048 (2015).
\bibitem{Wang2015} F. Wang, S. A. Kivelson, and D.-H. Lee, Nat. Phys. {\bf 11}, 959 (2015).
\bibitem{Shao2016} H. Shao, W. Guo, and A. W. Sandvik, Science {\bf 352}, 213 (2016).
\bibitem{Nahum2015B} A. Nahum, P. Serna, J. T. Chalker, M. Ortu\~{n}o, and A. M. Somoza, Phys. Rev. Lett. {\bf 115}, 267203 (2015).
\bibitem{Sato2017} T. Sato, M. Hohenadler, and F. F. Assaad, Phys. Rev. Lett. {\bf 119}, 197203 (2017).
\bibitem{Sreejith2019} G. J. Sreejith, S. Powell, and A. Nahum, Phys. Rev. Lett. {\bf 122}, 080601 (2019).
\bibitem{Yao2019} Z.-X. Li, S.-K. Jian, and H. Yao, arXiv:1904.10975

\bibitem{Landau} L. D. Landau and E. M. Lifshitz, {\it Statistical Physics} (Butterworth-Heinemann, 1999).

\bibitem{Li2017} Z.-X. Li, Y.-F. Jiang, S.-K. Jian, and H. Yao, Nat. Commun. {\bf 8}, 314 (2017).
\bibitem{Scherer2016} M. M. Scherer and I. F. Herbut, Phys. Rev. B {\bf 94}, 205136 (2016).
\bibitem{Classen2017} L. Classen, I. F. Herbut and M. M. Scherer, Phys. Rev. B {\bf 96}, 115132 (2017).
\bibitem{Jian2017A} S.-K. Jian and H. Yao, Phys. Rev. B {\bf 96}, 195162 (2017).
\bibitem{Jian2017B} S.-K. Jian and H. Yao, Phys. Rev. B {\bf 96}, 155112 (2017).
\bibitem{Torres2018} E. Torres, L. Classen, I. F. Herbut and M. M. Scherer, Phys. Rev. B {\bf 97}, 125137 (2018).
\bibitem{Roy2019} B. Roy and V. Juri\v{c}i\'{c}, Phys. Rev. B {\bf 99}, 241103 (2019).
\bibitem{Yin2020} S. Yin and Z. Y. Zuo, Phys. Rev. B {\bf 101}, 155136 (2020).
\bibitem{Devonshire} A. F. Devonshire, Philos. Mag. {\bf 40}, 1040 (1949).
\bibitem{Stephanov1995} M. A. Stephanov, Phys. Rev. D {\bf 52}, 3746 (1995).
\bibitem{Wessel2016} S. Hesselmann and S. Wessel, Phys. Rev. B {\bf 93}, 155157 (2016).

\bibitem{supmat} See Supplemental Material for details.

\bibitem{Boyackreview} R. Boyack, H. Yerzhakov, and J. Maciejko, arXiv: 2004.09414.
\bibitem{Mitraprl2011} A. Mitra and T. Giamarchi, Phys. Rev. Lett. {\bf 107}, 150602 (2011).
\bibitem{Mitraprl2012} A. Mitra, Phys. Rev. Lett. {\bf 109}, 260601 (2012).


\bibitem{Srednicki2015} K. R. Fratus and M. Srednicki, Phys. Rev. E {\bf 92}, 040103(R) (2015).


\bibitem{Greif2015}D. Greif, G. Jotzu, M. Messer, R. Desbuquois, and T. Esslinger, Phys. Rev. Lett. {\bf 115}, 260401 (2015).
\bibitem{Bloch2017} C. Gross and I. Bloch, Science {\bf 357}, 995 (2017).
\bibitem{Weiss2006} T. Kinoshita, T. Wenger, and D. S. Weiss, Nature {\bf 440}, 900 (2006).
\bibitem{Hofferberth2007} S. Hofferberth, I. Lesanovsky, B. Fischer, T. Schumm, and J. Schmiedmayer, Nature {\bf 449}, 324 (2007).
\bibitem{Erne} S. Erne, R. B\"{u}cker, T. Gasenzer, J. Berges, and J. Schmiedmayer, Nature (London) {\bf 563}, 225 (2018).
\bibitem{Oberthaler} M. Pr\"{u}fer, P. Kunkel, H. Strobel, S. Lannig, D. Linnemann, C. Schmied, J. Berges, T. Gasenzer, and M. K. Oberthaler, Nature (London) {\bf 563}, 217 (2018).
\bibitem{Nicklas} E. Nicklas, M. Karl, M. H\"{o}fer, A. Johnson, W. Muessel, H. Strobel, and J. Tomkovi\v{c}, T. Gasenzer, and M. K. Oberthaler, Phys. Rev. Lett. {\bf 115}, 245301 (2015).

\bibitem{Jakubczyk} P. Jakubczyk, Phys. Rev. B {\bf 79}, 125115 (2009).
\bibitem{Jakubczyk2009} P. Jakubczyk, W. Metzner, and H. Yamase, Phys. Rev. Lett. {\bf 103}, 220602 (2009).

\end{thebibliography}

\begin{thebibliography}{9}
\bibitem{Mitra2015S} A. Chiocchetta, M. Tavora, A. Gambassi, and A. Mitra, Phys. Rev. B {\bf 91}, 220302(R) (2015).
\bibitem{Mitra20151S} A. Maraga, A. Chiocchetta, A. Mitra, and A. Gambassi, Phys. Rev. E {\bf 92}, 042151 (2015).
\bibitem{Mitra2016S} A. Chiocchetta, M. Tavora, A. Gambassi, and A. Mitra, Phys. Rev. B {\bf 94}, 134311 (2016).
\bibitem{Marino2017S} A. Chiocchetta, A. Gambassi, S. Diehl, and J. Marino, Phys. Rev. Lett. {\bf 118}, 135701 (2017).
\bibitem{Swingle2019S} S.-K. Jian, S. Yin, and B. Swingle, Phys. Rev. Lett. {\bf 123}, 170606 (2019).
\bibitem{Yin2020S} S. Yin and Z. Y. Zuo, Phys. Rev. B {\bf 101}, 155136 (2020).


\end{thebibliography}
\end{document}